\documentclass[10pt]{iopartmod}
\usepackage{amssymb}
\usepackage{amsmath}
\usepackage{makeidx}
\usepackage[dvips]{graphicx}

\input{tcilatex}

\begin{document}

\title{Magnetic Faraday rotation in lossy photonic structures}
\author{A. Figotin and I. Vitebskiy}

\begin{abstract}
Magnetic Faraday rotation is widely used in optics and MW. In uniform
magneto-optical materials, this effect is very weak. One way to enhance it
is to incorporate the magnetic material into a high-Q optical resonator. One
problem with magneto-optical resonators is that along with Faraday rotation,
the absorption and linear birefringence can also increase dramatically,
compromising the device performance.  Another problem is strong ellipticity
of the output light. We discuss how the above problems can be addressed in
the cases of optical microcavities and a slow wave resonators. We show that
a slow wave resonator has a fundamental advantage when it comes to Faraday
rotation enhancement in lossy magnetic materials.
\end{abstract}

\maketitle

\section{Introduction}

Magnetic materials play a crucial role in optics. They are essential in
numerous non-reciprocal devices such as optical isolators, circulators,
phase shifters, etc. A well-known example of nonreciprocal effects is
magnetic Faraday rotation related to nonreciprocal circular birefringence.
Nonreciprocal effects only occur in magnetically ordered materials, such as
ferromagnets and ferrites, or in the presence of bias magnetic field \cite%
{LLEM,Gurev}. At optical frequencies, all nonreciprocal effects are very
weak, and can be further obscured by absorption, linear and/or form
birefringence, etc. A\ way to enhance a weak Faraday rotation is to
incorporate the magneto-optical material into a resonator, which can be a
complex nanophotonic structure with feature sizes comparable to the light
wavelength \cite{MPC Inoue06,Lyubch,MPC
Inoue99,Levy07,Levy06,Grishin07,Grishin04,Vinogr}. An intuitive explanation
for the resonance enlacement invokes a simple idea that in a high-Q optical
resonator filled with magneto-optical material, each individual photon
resides much longer compared to the same piece of magnetic material taken
out of the resonator. Since the nonreciprocal circular birefringence is
independent of the direction of light propagation, one can assume that the
magnitude of Faraday rotation is proportional to the photon residence time
in the magnetic material. With certain reservations, the above assumption
does provide a hand-waving explanation of the resonance enhancement of
magnetic Faraday rotation, as well as many other light-matter interactions. 

Resonance conditions can indeed result in a significant enhancement of
nonreciprocal effects, which in our case is a desirable outcome. On the
other hand, the same resonance conditions can also enhance absorption and
linear birefringence in the same magnetic material, which would be
undesirable. Indeed, linear and/or form birefringence, if present, can
significantly suppress the Faraday rotation, or any other manifestation of
nonreciprocal circular birefringence. Even more damaging can be absorption.
In uniform magneto-optical materials, the absorption contributes to the
ellipticity of propagating electromagnetic wave by causing circular
dichroism. In low-loss uniform magnetic materials those effects are
insignificant. Under the resonance condition, though, the role of absorption
can change dramatically. Firstly, the enhanced absorption reduces the
intensity of light transmitted through the optical resonator. Secondly, even
moderate absorption can lower the Q-factor of the resonance by several
orders of magnitude and, thereby, significantly compromise its performance
as Faraday rotation enhancer. Finally, enhanced absorption, along with
spatial nonuniformity, contributes to deviation of the transmitted light
polarization from linear, making it difficult to measure the amount of
Faraday rotation.

We explore the idea of composite magneto-photonic structures having enhanced
nonreciprocal characteristics associated with magnetism but, at the same
time, significantly reducing the light absorption. In other words, we want
to enhance the useful characteristics of a particular magnetic material,
while drastically reducing its contribution to the energy dissipation. The
possibility of appreciable enhancement of Faraday rotation or other
nonreciprocal effects is particularly important at infrared and optical
frequencies, where all light-matter interactions are very weak. In those
cases, the use of photonic structures instead of uniform magnetic materials
can also dramatically reduce the size of the respective optical components,
without compromising their performance.

We also compare two qualitatively different approaches to resonance
enhancement of light-matter interactions. The first one is based on a
magnetic microcavity sandwiched between a pair of Bragg reflectors, as shown
in Fig. \ref{MC}. The second approach is based on a slow wave resonance in a
magnetic photonic crystal, an example of which is shown in Fig. \ref{AFStack}%
. In either case, one can simultaneously enhance the useful characteristics
of a particular magnetic material, while reducing its contribution to the
energy dissipation. Yet, the above two approaches are qualitatively
different, and which one is preferable depends on specific circumstances.
For instance, if the absorption of light by the magnetic material is an
issue, the slow wave resonance is definitely preferable. Otherwise, if the
light absorption is insignificant and the only goal is to enhance the
magnetic Faraday rotation, then the microcavity resonance can be a better
choice.

\section{Absorption suppression in composite structures}

How is it possible to enhance Faraday rotation produced by the lossy
magnetic component of composite structure, while reducing the losses caused
by the same magnetic material? Following \cite{PRB08}, we can use the fact
that the absorption and the useful functionality of the particular magnetic
material are related to different components of its permittivity and/or
permeability tensors $\hat{\varepsilon}$ and $\hat{\mu}$. Specifically, the
absorption is determined by the anti-Hermitian parts $\hat{\varepsilon}%
^{\prime \prime }$ and $\hat{\mu}^{\prime \prime }$ the permittivity and
permeability tensors%
\begin{equation}
\hat{\varepsilon}^{\prime \prime }=-\frac{i}{2}\left( \hat{\varepsilon}-\hat{%
\varepsilon}^{\dag }\right) ,\ \hat{\mu}^{\prime \prime }=-\frac{i}{2}\left( 
\hat{\mu}-\hat{\mu}^{\dag }\right) ,  \label{abs}
\end{equation}%
while the nonreciprocal circular birefringence responsible for the Faraday
rotation is determined by the Hermitian skew-symmetric parts of the
respective tensors%
\begin{equation}
\hat{\varepsilon}_{a}=\frac{i}{2}\func{Im}\left( \hat{\varepsilon}+\hat{%
\varepsilon}^{\dag }\right) ,\ \hat{\mu}_{a}=\frac{i}{2}\func{Im}\left( \hat{%
\mu}+\hat{\mu}^{\dag }\right) ,  \label{skews}
\end{equation}
where $\dag $ denotes Hermitian conjugate. The relations (\ref{abs}) and (%
\ref{skews}) suggest that the rate of energy absorption by the lossy
magnetic material can be functionally different from its useful
functionality (nonreciprocal circular birefringence in our case). Such a
difference allows us to adjust the physical and geometric characteristics of
the periodic structure so that the electromagnetic field distribution inside
the photonic structure suppresses the energy dissipation by the lossy
magnetic component, while even enhancing its useful functionality. The way
to address the problem essentially depends on the following factors.

\begin{enumerate}
\item The physical mechanism of Faraday rotation.

\item The dominant physical mechanism of absorption. For instance, energy
dissipation caused by electric conductivity requires a different approach,
compared to the situation where the losses are associated with the dynamics
of magnetic domains, or some other physical mechanisms. In each individual
case, the structure of the anti-Hermitian part (\ref{abs}) of the
permittivity and/or permeability tensors can be different, and so can be the
optimal configuration of the composite material.

\item The frequency range of interest. A given photonic structure can
dramatically enhance Faraday rotation at some frequencies, while sharply
reducing it at different frequencies. The same is true with absorption,
which can be either suppressed, or enhanced, depending on the frequency
range. 
\end{enumerate}

Since our goal is to enhance Faraday rotation while reducing absorption, the
same photonic structure can be either effective or counterproductive,
depending on the frequency range and the dominant physical mechanism of
electromagnetic energy dissipation. Fortunately, in some important cases,
the photonic structure can be engineered in such a way that it only enhances
the useful light-matter interaction, while limiting or even suppressing the
absorption. Usually, it can be done if the useful functionality and the
absorption are associated with different components of electromagnetic
field. An impressive example of the kind is considered in \cite{PRB08},
where a simple layered structure provides significant enhancement of Faraday
rotation produced by a lossy magnetic component, while dramatically reducing
absorption caused by the same magneto-optical material.

Under what circumstances can we not only suppress the absorption but also
have the size of the periodic composite structure much smaller than that of
the uniform (magnetic) slab with similar performance? When considering this
question we should keep in mind that within the framework of the photonic
approach the characteristic length $L$ of the the structural components is
always comparable to that of the electromagnetic wavelength in the medium.
Therefore, for a given frequency range and for a given set of the
constitutive materials, we cannot significantly change the length $L$. Nor
can we substantially reduce the number $N$\textit{\ }of unit cells of the
periodic structure without loosing all the effects of coherent interference.
All we can achieve by adjusting the configuration of the periodic array
comprising as few as several periods is to suppress the losses and/or to
enhance the Faraday rotation. The real question is: what is the thickness $%
D_{U}$ of the uniform slab producing Faraday rotation comparable to that of
the optimized photonic structure? Indeed, if such a uniform slab turns out
to be much thicker than the layered structure, then we can claim that not
only the periodic array dramatically reduces the losses, but it also has
much smaller dimensions. The latter is only possible if the thickness $D_{U}$
of the uniform slab with desired functionality is much greater than the
electromagnetic wavelength in the medium. Otherwise, all we can achieve by
introducing periodic inhomogeniety would be a reduction of losses. At
optical frequencies, due to the weakness of light-matter interactions, the
thickness of the uniform slab producing significant Faraday rotation is
indeed much greater than the light wavelength. Therefore, in optics we can
simultaneously suppress the losses, while reducing the size of the
nonreciprocal optical device.

\TEXTsymbol{>}\TEXTsymbol{>}\TEXTsymbol{>}\TEXTsymbol{>}\TEXTsymbol{>}%
\TEXTsymbol{>}\TEXTsymbol{>}\TEXTsymbol{>}\TEXTsymbol{>}\TEXTsymbol{>}%
\TEXTsymbol{>}\TEXTsymbol{>}\TEXTsymbol{>}\TEXTsymbol{>}

If the resonance Q-factor is high enough, the acquired ellipticity becomes
so significant that the very term "Faraday rotation" becomes irrelevant.
Indeed, one cannot assign a meaningful rotation angle to a wave with nearly
circular polarization. The above circumstance, though, does not diminish the
practical importance of the nonreciprocal effect, which now reduces to the
conversion of linear polarization of the incident wave to nearly circular
polarization of transmitted and/or reflected waves.

\section{Notations, definitions, and physical assumptions}

\subsection{Transverse electromagnetic waves in stratified media}

Our analysis is based on the time-harmonic Maxwell equations%
\begin{equation}
\nabla\times\vec{E}\left( \vec{r}\right) =i\frac{\omega}{c}\hat{\mu}\left( 
\vec{r}\right) \vec{H}\left( \vec{r}\right) ,\;\nabla\times\vec{H}\left( 
\vec{r}\right) =-i\frac{\omega}{c}\hat{\varepsilon}\left( \vec{r}\right) 
\vec{E}\left( \vec{r}\right) ,  \label{THME}
\end{equation}
where the second rank tensors $\hat{\varepsilon}\left( \vec{r}\right) $ and $%
\hat{\mu}\left( \vec{r}\right) $ are coordinate dependent. In a stratified
medium 
\begin{equation*}
\hat{\varepsilon}\left( \vec{r}\right) =\hat{\varepsilon}\left( z\right) ,%
\hat{\mu}\left( \vec{r}\right) =\hat{\mu}\left( z\right) ,
\end{equation*}
where the Cartesian coordinate $z$ is normal to the layers. We also assume
that the dielectric permittivity and magnetic permeability tensors in each
layer has the following form%
\begin{equation}
\hat{\varepsilon}=\left[ 
\begin{array}{ccc}
\varepsilon_{xx} & \varepsilon_{xy} & 0 \\ 
\varepsilon_{yx} & \varepsilon_{yy} & 0 \\ 
0 & 0 & \varepsilon_{zz}%
\end{array}
\right] ,~~\hat{\mu}=\left[ 
\begin{array}{ccc}
\mu_{xx} & \mu_{xy} & 0 \\ 
\mu_{yx} & \mu_{yy} & 0 \\ 
0 & 0 & \mu_{zz}%
\end{array}
\right] ,  \label{eps mu trnsv}
\end{equation}
in which case the layered structure support transverse electromagnetic waves
with%
\begin{equation}
\vec{E}\left( \vec{r}\right) =\vec{E}\left( z\right) \perp z,\ \vec {H}%
\left( \vec{r}\right) =\vec{H}\left( z\right) \perp z,  \label{EH trsv}
\end{equation}
propagating along the $z$ direction. The Maxwell equations (\ref{THME}) in
this case reduce to the following system of four ordinary differential
equations%
\begin{equation}
\frac{\partial}{\partial z}\Psi\left( z\right) =i\frac{\omega}{c}M\left(
z\right) \Psi\left( z\right) ,  \label{ME4}
\end{equation}
\ where%
\begin{equation}
\Psi\left( z\right) =\left[ 
\begin{array}{c}
E_{x}\left( z\right) \\ 
E_{y}\left( z\right) \\ 
H_{x}\left( z\right) \\ 
H_{y}\left( z\right)%
\end{array}
\right] ,  \label{Psi}
\end{equation}
and%
\begin{equation}
M\left( z\right) =\left[ 
\begin{array}{cccc}
0 & 0 & \mu_{xy}^{\ast} & \mu_{yy} \\ 
0 & 0 & -\mu_{xx} & -\mu_{xy} \\ 
-\varepsilon_{xy}^{\ast} & -\varepsilon_{yy} & 0 & 0 \\ 
\varepsilon_{xx} & \varepsilon_{xy} & 0 & 0%
\end{array}
\right] .  \label{M trsv}
\end{equation}
The $4\times4$ matrix $M\left( z\right) $ is referred to as the (reduced)
Maxwell operator.

Solutions for the reduced time-harmonic Maxwell equation (\ref{ME4}) can be
presented in the following form%
\begin{equation}
\Psi\left( z\right) =T\left( z,z_{0}\right) \Psi\left( z_{0}\right) ,
\label{T(zz0)}
\end{equation}
where the $4\times4$ matrix $T\left( z,z_{0}\right) $ is the \emph{transfer
matrix}. The transfer matrix (\ref{T(zz0)}) uniquely relates the values of
electromagnetic field (\ref{Psi}) at any two points $z$ and $z_{0}$ of the
stratified medium.

In a uniform medium, the Maxwell operator $M$ in (\ref{M trsv}) is
independent of $z$. In this case, the transfer matrix $T\left(
z,z_{0}\right) $ can be explicitly expressed in terms of the respective
Maxwell operator $M$%
\begin{equation}
T\left( z,z_{0}\right) =\exp\left[ i\frac{\omega}{c}\left( z-z_{0}\right) M%
\right] .  \label{T hmg}
\end{equation}
In particular, the transfer matrix of an individual uniform layer $m$ is%
\begin{equation}
T_{m}=\exp\left( i\frac{\omega}{c}z_{m}M_{m}\right) ,  \label{Tm}
\end{equation}
where $z_{m}$ is the thickness of the $m$-th layer.

The transfer matrix $T_{S}$ of an arbitrary stack of layers is a sequential
product of the transfer matrices $T_{m}$ of the constituent layers%
\begin{equation}
T_{S}=\prod_{m}T_{m}.  \label{TS}
\end{equation}

In the following subsection we specify the form of the material tensors (\ref%
{eps mu trnsv}), which determine the transfer matrices of the individual
layers and the entire periodic structure. In this paper, we use the same
notations as in our previous publication \cite{PRE01,PRB03,PRB08} related to
magnetic layered structures.

\subsection{Permittivity and permeability tensors of the layers}

We assume that the permittivity and permeability tensors of individual
layers have the following form%
\begin{equation}
\hat{\varepsilon}=\left[ 
\begin{array}{ccc}
\varepsilon+\delta & i\alpha & 0 \\ 
-i\alpha & \varepsilon-\delta & 0 \\ 
0 & 0 & \varepsilon_{zz}%
\end{array}
\right] ,~~\hat{\mu}=1,  \label{eps AF, mu AF}
\end{equation}
where $\alpha$ is responsible for nonreciprocal circular birefringence and $%
\delta$ describes linear birefringence. In a lossless medium, the physical
quantities $\varepsilon$, $\alpha$, and $\delta$ are real. If the direction
of magnetization is changed for the opposite, the parameters $\alpha$ also
changes its sign and so will the sense of Faraday rotation \cite{LLEM,Gurev}%
. The absorption, is accounted for by allowing $\varepsilon$, $\alpha$, and $%
\delta$ to be complex.

Substitution of (\ref{eps AF, mu AF}) into (\ref{M trsv}) yields the
following expression for the Maxwell operator%
\begin{equation}
M=\left[ 
\begin{array}{cccc}
0 & 0 & 0 & 1 \\ 
0 & 0 & -1 & 0 \\ 
i\alpha & -\varepsilon+\delta & 0 & 0 \\ 
\varepsilon+\delta & i\alpha & 0 & 0%
\end{array}
\right] .  \label{M AF}
\end{equation}
The respective four eigenvectors are 
\begin{equation}
\left\{ 
\begin{array}{c}
1 \\ 
-ir_{1} \\ 
in_{1}r_{1} \\ 
n_{1}%
\end{array}
\right\} \leftrightarrow n_{1},\ \left\{ 
\begin{array}{c}
1 \\ 
-ir_{1} \\ 
-in_{1}r_{1} \\ 
-n_{1}%
\end{array}
\right\} \leftrightarrow-n_{1},\ \left\{ 
\begin{array}{c}
-ir_{2} \\ 
1 \\ 
-n_{2} \\ 
-in_{2}r_{2}%
\end{array}
\right\} \leftrightarrow n_{2},\ \left\{ 
\begin{array}{c}
-ir_{2} \\ 
1 \\ 
n_{2} \\ 
in_{2}r_{2}%
\end{array}
\right\} \leftrightarrow-n_{2}.  \label{EV AF}
\end{equation}
$\allowbreak\allowbreak$where%
\begin{equation}
n_{1}=\sqrt{\varepsilon+\sqrt{\delta^{2}+\alpha^{2}}},\ \ n_{2}=\sqrt {%
\varepsilon-\sqrt{\delta^{2}+\alpha^{2}}},  \label{n_1, n_2}
\end{equation}%
\begin{equation}
r_{1}=\frac{\alpha}{\sqrt{\delta^{2}+\alpha^{2}}+\delta},\ r_{2}=\frac {%
\sqrt{\delta^{2}+\alpha^{2}}-\delta}{\alpha},  \label{r_1, r_2}
\end{equation}
Compared to \cite{Levy07}, we use slightly different notations.

The explicit expression for the transfer matrix $\hat{T}(A)$ of a single
uniform layer of thickness $A$ is%
\begin{equation}
\hat{T}(A)=\hat{W}\left( A\right) \hat{W}^{-1}(0),  \label{TG}
\end{equation}
where%
\begin{equation}
\hat{W}\left( A\right) =\left[ 
\begin{array}{llll}
e^{i\phi_{1}} & e^{-i\phi_{1}} & -ir_{2}e^{i\phi_{2}} & -ir_{2}e^{-i\phi_{2}}
\\ 
-ir_{1}e^{i\phi_{1}} & -ir_{1}e^{-i\phi_{1}} & e^{i\phi_{2}} & e^{-i\phi_{2}}
\\ 
ir_{1}n_{1}e^{i\phi_{1}} & -ir_{1}n_{1}e^{-i\phi_{1}} & -n_{2}e^{i\phi_{2}}
& n_{2}e^{-i\phi_{2}} \\ 
n_{1}e^{i\phi_{1}} & -n_{1}e^{-i\phi_{1}} & -ir_{2}n_{2}e^{i\phi_{2}} & 
ir_{2}n_{2}e^{-i\phi_{2}}%
\end{array}
\right] ,  \label{WG}
\end{equation}
and%
\begin{equation*}
\phi_{1}=\frac{\omega}{c}An_{1},\ \ \phi_{2}=\frac{\omega}{c}An_{2}.
\end{equation*}

The eigenvectors (\ref{EV AF}) correspond to elliptically polarized states.
There are two important particular cases corresponding to linearly and
circularly polarized eigenmodes, respectively.

\subsubsection{Non-magnetic medium with linear birefringence}

In the case of a non-magnetic medium%
\begin{equation}
\alpha=0,\ r_{1}=0,\ r_{2}=0.  \label{alpha=0}
\end{equation}
The respective eigenmodes are linearly polarized%
\begin{equation}
\left\{ 
\begin{array}{c}
1 \\ 
0 \\ 
0 \\ 
n_{1}%
\end{array}
\right\} \leftrightarrow n_{1},\ \left\{ 
\begin{array}{c}
1 \\ 
0 \\ 
0 \\ 
-n_{1}%
\end{array}
\right\} \leftrightarrow-n_{1},\ \left\{ 
\begin{array}{c}
0 \\ 
1 \\ 
-n_{2} \\ 
0%
\end{array}
\right\} \leftrightarrow n_{2},\ \left\{ 
\begin{array}{c}
0 \\ 
1 \\ 
n_{2} \\ 
0%
\end{array}
\right\} \leftrightarrow-n_{2}.  \label{EV A0}
\end{equation}
where%
\begin{equation*}
n_{1}=\sqrt{\varepsilon+\delta},\ \ n_{2}=\sqrt{\varepsilon-\delta}.
\end{equation*}

\subsubsection{Magnetic medium with circular birefringence}

Another important limiting case corresponds to a uniaxial magnetic medium
with%
\begin{equation}
\delta=0,\ r_{1}=1,\ r_{2}=1.  \label{delta=0}
\end{equation}
The respective eigenmodes are circularly polarized 
\begin{equation}
\left\{ 
\begin{array}{c}
1 \\ 
-i \\ 
in_{1} \\ 
n_{1}%
\end{array}
\right\} \leftrightarrow n_{1},\ \left\{ 
\begin{array}{c}
1 \\ 
-i \\ 
-in_{1} \\ 
-n_{1}%
\end{array}
\right\} \leftrightarrow-n_{1},\ \left\{ 
\begin{array}{c}
-i \\ 
1 \\ 
-n_{2} \\ 
-in_{2}%
\end{array}
\right\} \leftrightarrow n_{2},\ \left\{ 
\begin{array}{c}
-i \\ 
1 \\ 
n_{2} \\ 
in_{2}%
\end{array}
\right\} \leftrightarrow-n_{2}.  \label{EV F}
\end{equation}
where%
\begin{equation*}
n_{1}=\sqrt{\varepsilon+\alpha},\ \ n_{2}=\sqrt{\varepsilon-\alpha}.
\end{equation*}

\subsection{Numerical values of material tensors}

Our objectives include two distinct problems associated with Faraday
rotation enhancement.

One problem can be caused by the presence of linear birefringence described
by the parameter $\delta$ in (\ref{eps AF, mu AF}). Linear birefringence $%
\delta$ competes with circular birefringence $\alpha$. At optical
frequencies, the former can easily prevail and virtually annihilate any
manifestations of nonreciprocal circular birefringence. If linear
birefringence occurs in magnetic F layers in Fig. \ref{AFStack}, it can be
offset by linear birefringence in the alternating dielectric A layers.
Similarly, in the case of a magnetic resonance cavity in Fig. \ref{MC}, the
destructive effect of the linear birefringence in the magnetic D layer can
be offset by linear birefringence in layers constituting the Bragg
reflectors. In either case, the cancellation of linear birefringence of the
magnetic layers only takes place at one particular frequency. Therefore, the
layered structure should be designed so that this particular frequency
coincides with the operational resonance frequency of the composite
structure. The detailed discussion on the effect of linear birefringence and
ways to deal with it will be presented elsewhere.

In the rest of the paper we will focus on the problem associated with
absorption. This problem is unrelated to the presence or absence of linear
birefringence and, therefore, can be handled separately. For this reason, in
our numerical simulation we can set $\delta=0$ and use the following
expressions for the dielectric permittivity tenors of the magnetic F-layers
and dielectric A-layers in Fig. \ref{AFStack} 
\begin{equation}
\hat{\varepsilon}_{F}=\left[ 
\begin{array}{ccc}
\varepsilon_{F}+i\gamma & i\alpha & 0 \\ 
-i\alpha & \varepsilon_{F}+i\gamma & 0 \\ 
0 & 0 & \varepsilon_{3}%
\end{array}
\right] ,  \label{eps_F}
\end{equation}%
\begin{equation}
\hat{\varepsilon}_{A}=\left[ 
\begin{array}{ccc}
\varepsilon_{A} & 0 & 0 \\ 
0 & \varepsilon_{A} & 0 \\ 
0 & 0 & \varepsilon_{A}%
\end{array}
\right] ,  \label{eps_A}
\end{equation}
where $\varepsilon_{F}$, $\varepsilon_{A}$, and $\gamma$ are real. Parameter 
$\gamma$ describes absorption of the magnetic material.

In the case of photonic cavity in Fig. \ref{MC} we use similar material
parameters. The permittivity tensor of the magnetic D-layer is the same as
that of the magnetic F-layers in Fig. \ref{AFStack}%
\begin{equation}
\hat{\varepsilon}_{D}=\hat{\varepsilon}_{F}  \label{eps_D}
\end{equation}
$\hat{\varepsilon}_{F}$ is defined in (\ref{eps_F}). The permittivity
tensors of the alternating dielectric layers A and B constituting the Bragg
reflectors in Fig. \ref{MC} are chosen as follows%
\begin{equation}
\hat{\varepsilon}_{B}=\left[ 
\begin{array}{ccc}
\varepsilon_{B} & o & 0 \\ 
0 & \varepsilon_{B} & 0 \\ 
0 & 0 & \varepsilon_{B}%
\end{array}
\right] ,\ \ \hat{\varepsilon}_{C}=\left[ 
\begin{array}{ccc}
\varepsilon_{C} & 0 & 0 \\ 
0 & \varepsilon_{C} & 0 \\ 
0 & 0 & \varepsilon_{C}%
\end{array}
\right] .  \label{eps_A, eps_B}
\end{equation}

In either case, only the magnetic layers F or D are responsible for
absorption, which is a realistic assumption.

In the case of periodic stack in Fig. \ref{AFStack} we use the following
numerical values of the diagonal components of the permittivity tensors%
\begin{equation}
\varepsilon_{F}=5.37,\ \ \varepsilon_{A}=2.1.  \label{Num PS}
\end{equation}
Similar values are used in the case of photonic microcavity in Fig. \ref{MC}%
\begin{equation*}
\varepsilon_{D}=\varepsilon_{C}=5.37,\ \ \varepsilon_{B}=2.1.
\end{equation*}
The numerical values of the gyrotropic parameter $\alpha$, as well as the
absorption coefficient $\gamma$ of the magnetic layers F and D, remain
variable. We also tried different layer thicknesses $d_{A}$,$\ d_{F}$,$\
d_{B}$, $d_{C}$, and $\ d_{D}$. But in this paper we only include the
results corresponding to the following numerical values%
\begin{equation}
d_{A}=d_{C}=0.8L,\ d_{F}=d_{C}=0.2L,\ d_{D}=0.4L,  \label{Num d}
\end{equation}
where $L$ is the length of a unit cell of the periodic array%
\begin{equation*}
L=d_{F}+d_{A}=d_{B}+d_{C}.
\end{equation*}
The thickness $d_{D}$ of the defect layer in Fig. \ref{MC} is chosen so that
the frequency of the defect mode falls in the middle of the lowest photonic
band gap of Bragg reflectors.

\subsection{Scattering problem for magnetic layered structure}

In all cases, the incident wave $\Psi_{I}$ propagates along the $z$
direction normal to the layers. Unless otherwise explicitly stated, the
incident wave polarization is linear with $\vec{E}_{I}\parallel x$. Due to
the nonreciprocal circular birefringence of the magnetic material, the
transmitted and reflected waves $\Psi_{P}$ and $\Psi_{R}$ will be
elliptically polarized with the ellipse axes being at an angle with the $x$
direction.

The transmitted and reflected waves, as well the electromagnetic field
distribution inside the layered structure, are found using the transfer
matrix approach. Let us assume that the left-hand and the right-hand
boundaries of a layered array are located at $z=0$ and $a=d$, respectively.
According to (\ref{T(zz0)}) and (\ref{TS}), the incident, transmitted, and
reflected waves are related as follows%
\begin{equation}
\Psi_{P}(d)=T_{S}\left( \Psi_{I}(0)+\Psi_{R}(0)\right) .  \label{BC}
\end{equation}
Knowing the incident wave $\Psi_{I}$ and the transfer matrix $T_{S}$ of the
entire layered structure and assuming, we can solve the system (\ref{BC}) of
four linear equations and, thereby, find the reflected and transmitted
waves. Similarly, using the relation (\ref{T(zz0)}), we can also find the
field distribution inside the layered structure.

The transmission and reflection coefficients of the slab (either uniform, or
layered) are defined as follows%
\begin{equation}
t=\frac{S_{P}}{S_{I}},\ r=-\frac{S_{R}}{S_{I}},  \label{t,r}
\end{equation}
where $S_{I}$, $S_{P}$, and $S_{R}$ are the Poynting vectors of the
incident, transmitted, and reflected waves, respectively. The slab
absorption is 
\begin{equation}
a=1-t-r.  \label{a}
\end{equation}

If the incident wave polarization is linear, the coefficients $t$, $r$, and $%
a$ are independent of the orientation of vector $\vec{E}_{I}$ in the $x-y$
plane, because for now, we neglect the linear birefringence $\delta$. Due to
nonreciprocal circular birefringence, the polarization of the transmitted
and reflected waves will always be elliptic.

By contrast, if the incident wave polarization is circular, the coefficients 
$t$, $r$, and $a$ depend on the sense of circular polarization. The
polarization of the transmitted and reflected waves in this case will be
circular with the same sense of rotation as that of the incident wave.

The effect of nonreciprocal circular birefringence on transmitted wave can
be quantified by the following expression%
\begin{equation}
\Delta\Psi_{P}=\frac{1}{2}\left[ \left( \Psi_{P}\right) _{\alpha}-\left(
\Psi_{P}\right) _{-\alpha}\right]  \label{NR Diff}
\end{equation}
where $\left( \Psi_{P}\right) _{\alpha}$ and $\left( \Psi_{P}\right)
_{-\alpha}$ respectively correspond to the wave transmitted through the
original periodic structure and through the same structure but with the
opposite sign of circular birefringence parameter $\alpha$. If the incident
wave polarization is linear with $\vec{E}_{I}\parallel x$, the vector-column
(\ref{NR Diff}) has the following simple structure%
\begin{equation*}
\Delta\Psi_{P}=\left( \vec{E}_{P}\right) _{y}\left[ 
\begin{array}{c}
0 \\ 
1 \\ 
1 \\ 
0%
\end{array}
\right] ,
\end{equation*}
implying that the $y$ component $\left( \vec{E}_{P}\right) _{y}$ of the
transmitted wave has "purely" nonreciprocal origin and, therefore, can used
to characterize the magnitude of nonreciprocal circular birefringence on
transmitted wave. Indeed, in the absence of magnetism, the parameter $\alpha$
in (\ref{eps AF, mu AF}), (\ref{eps_F}), and \ref{eps_D}) vanishes and the
transmitted wave is linearly polarized with $\vec{E}_{P}\parallel x$. The
above statement follows directly from symmetry consideration and remains
valid even in the presence of linear birefringence $\delta$ in (\ref{eps AF,
mu AF}). Further in this paper will use the ratio%
\begin{equation}
\rho=\frac{\left( E_{P}\right) _{y}}{\left( E_{I}\right) _{x}},\text{ \
where \ }\left\vert \rho\right\vert <1.  \label{rho}
\end{equation}
to characterize the effect circular birefringence on transmitted wave.

Generally, the transmitted wave polarization in the situation in Figs. \ref%
{AFStack} and \ref{MC} is elliptical, rather than linear. Therefore, the
quantity $\rho$ in (\ref{rho}) is not literally the sine of the Faraday
rotation angle. Let us elaborate on this point. The electromagnetic
eigenmodes of the layered structures in Figs. \ref{AFStack} and \ref{MC}
with permittivity tensors given in (\ref{eps_F}) through (\ref{eps_A, eps_B}%
) are all circularly polarized. This implies that if the polarization of the
incident wave is circular, the transmitted and reflected waves will also be
circularly polarized. On the other hand, due to the nonreciprocal (magnetic)
effects, the transmission/reflection coefficients for the right-hand
circular polarization are different from those for the left-hand circular
polarization. This is true regardless of the presence or absence of
absorption. Consider now \ a linearly polarized incident wave. It can be
viewed as a superposition of two circularly polarized waves with equal
amplitudes. Since the transmission/reflection coefficients for the
right-hand and left-hand circular polarizations are different, the
transmitted and reflected waves will be elliptically polarized. Such an
ellipticity develops both in the case of a uniform slab and in the case of a
layered stack, periodic or aperiodic, with or without absorption. Note,
though, that at optical frequencies, the dominant contribution to
ellipticity of the wave transmitted through a uniform slab is usually
determined by absorption, which is largely responsible for circular
dichroism. Without absorption, the ellipticity of the wave transmitted
through a uniform magnetic slab would be negligible. This might not be the
case for the layered structures in Figs. \ref{AFStack} and \ref{MC} at
frequencies of the respective transmission resonances. In these cases, the
ellipticity of transmitted and reflected waves can be significant even in
the absence of absorption. Moreover, if the Q-factor of the respective
resonance is high enough, the transmitted wave polarization becomes very
close to circular and, therefore, cannot be assigned any meaningful angle of
rotation. The numerical examples of the next section illustrate the above
statements.

To avoid confusion, note that a linear polarized wave propagating in a
uniform, lossless, unbounded, magnetic medium (\ref{eps_F}) will not develop
any ellipticity. Instead, it will display a pure Faraday rotation. But the
slab boundaries and the layer interfaces will produce some ellipticity even
in the case of lossless magnetic material. The absorption provides an
additional contribution to the ellipticity of transmitted and reflected
waves. The latter contribution is referred to as circular dichroism.

For simplicity, in further consideration we will often refer to the quantity 
$\rho$ in (\ref{rho}) as the amount of (nonreciprocal) Faraday rotation,
although, due to the ellipticity, it is not exactly the sine of the Faraday
rotation angle.

In all plots, the frequency $\omega$ and the Bloch wave number $k$ are
expressed in dimensionless units of $cL^{-1}$ and $L^{-1}$, respectively. In
our computations we use a transfer matrix approach identical to that
described in Ref. \cite{PRE01,PRB03}.

\section{Resonance enhancement of magnetic Faraday rotation}

\subsection{Cavity resonance: Lossless case}

Let us start with the resonance enhancement based on microcavity. The
magnetic layer D in Fig. \ref{MC} is sandwiched between two identical
periodic stacks playing the role of distributed Bragg reflectors. The
D-layer is also referred to as a defect layer, because without it, the
layered structure in Fig. \ref{MC} would be perfectly periodic. The
thickness of the defect layer is chosen so that the microcavity develops a
single resonance mode with the frequency lying in the middle of the lowest
photonic band gap of the adjacent periodic stacks. This resonance mode is
nearly localized in the vicinity of the magnetic D-layer.

A typical transmission spectrum of such a layered structure in the absence
of absorption is shown in Fig. \ref{t_FSDM_A0_G}. The stack transmission
develops a sharp peak at the defect mode frequency. The respective
transmission resonance is accompanied by a dramatic increase in field
amplitude in the vicinity of the magnetic D-layer. The large field amplitude
implies the enhancement of magnetic Faraday rotation produced by the
D-layer, as clearly seen in Fig. \ref{p_FSDM_A_G0}.

If the Q-factor of the microcavity exceeds certain value and/or if the
circular birefringence of the magnetic material of the D-layer is strong
enough, the resonance frequency of the defect mode splits into two, as shown
in Fig. \ref{p_FSDM_A_G0}(c) and (d). Each of the two resonances is
associated with left or right circular polarization. The transmitted light
will also display nearly perfect circular polarization with the opposite
sense of rotation for the twin resonances. Formally, the above nonreciprocal
effect cannot be classified as Faraday rotation, but it does not diminish
its practical value.

\subsection{Slow wave resonance: Lossless case}

The second approach to Faraday rotation enhancement is based on the
transmission band edge resonance in periodic stacks of magnetic layers
alternating with some other dielectric layers, as shown in Fig. \ref{AFStack}%
. A typical transmission spectrum of such a layered structure is shown in
Fig. \ref{t_PFS_a0_g}. The sharp peaks in transmission bands correspond to
transmission band edge resonances, also known as Fabry-Perot resonances. The
resonance frequencies are located close to a photonic band edge, where the
group velocity of the respective Bloch eigenmodes is very low. This is why
the transmission band edge resonances are referred to as slow wave
resonances. All resonance frequencies are located in transmission bands --
not in photonic band gaps, as in the case of a localized defect mode. The
resonance field distribution inside the periodic stack is close to a
standing wave composed of a pair of Bloch modes with equal and opposite
group velocities and nearly equal large amplitudes%
\begin{equation}
\Psi_{T}\left( z\right) =\Psi_{k}\left( z\right) +\Psi_{-k}\left( z\right) ,
\label{1sw}
\end{equation}
The left-hand and right-hand photonic crystal boundaries coincide with the
standing wave nodes, where the forward and backward Bloch components
interfere destructively to meet the boundary conditions. The most powerful
slow wave resonance corresponds to the transmission peak closest to the
respective photonic band edge, where the wave group velocity is lowest. At
resonance, the energy density distribution inside the periodic structure is
typical of a standing wave%
\begin{equation}
W\left( z\right) \propto W_{I}N^{2}\sin^{2}\left( \frac{\pi}{NL}z\right) ,
\label{W(z) RBE}
\end{equation}
where $W_{I}$ is the intensity of the incident light, $N$ is the total
number of unit cells (double layers) in the periodic stack in Fig. \ref%
{AFStack}.

Similarly to the case of magnetic cavity resonance, the large field
amplitude implies the enhancement of magnetic Faraday rotation produced by
magnetic F-layers, as demonstrated in Fig. \ref{p_PFS_a_g0}. Again, if the
Q-factor of the slow wave resonance exceeds certain value and/or if the
circular birefringence $\alpha$ of the magnetic material of the F-layers is
strong enough, each resonance frequency splits into two, as shown in Fig. %
\ref{t_PFS_a_g0_fwLx}. Each of the two twin resonances is associated with
left or right circular polarization. To demonstrate it, let us compare the
transmission dispersion in Fig. \ref{t_PFS_a_g0_fwLx}, where the incident
light polarization is linear, to the transmission dispersion in Figs. \ref%
{t_PFS_a_g0_fwLcp} and \ref{t_PFS_a_g0_fwLcn}, where the incident wave is
circularly polarized. One can see that the case in Fig. \ref{t_PFS_a_g0_fwLx}
of linearly polarized incident light reduces to a superposition of the cases
in Figs. \ref{t_PFS_a_g0_fwLcp} and \ref{t_PFS_a_g0_fwLcn} of two circularly
polarized incident waves with opposite sense of rotation.

\subsection{The role of absorption}

In the absence of absorption, the practical difference between cavity
resonance and slow wave resonance is not that obvious. But if the magnetic
material displays an appreciable absorption, the slow wave resonator is
definitely preferable. The physical reason for this is as follows.

In the case of a slow wave resonance, the reduction of the transmitted wave
energy is mainly associated with absorption. Indeed, although some fraction
of the incident light energy is reflected at the left-hand interface of the
periodic stack in Fig. \ref{AFStack}, this fraction remains limited even in
the case of strong absorption, as seen in Fig. \ref{ar_PFS_a0_g}(b). So, the
main source of the energy losses in a slow wave resonator is absorption,
which is a natural side effect of the Faraday rotation enhancement (some
important reservations can be found in \cite{PRB08}).

\begin{figure}[tbph]
\scalebox{0.8}{\includegraphics[viewport=-50 0 500 300,clip]{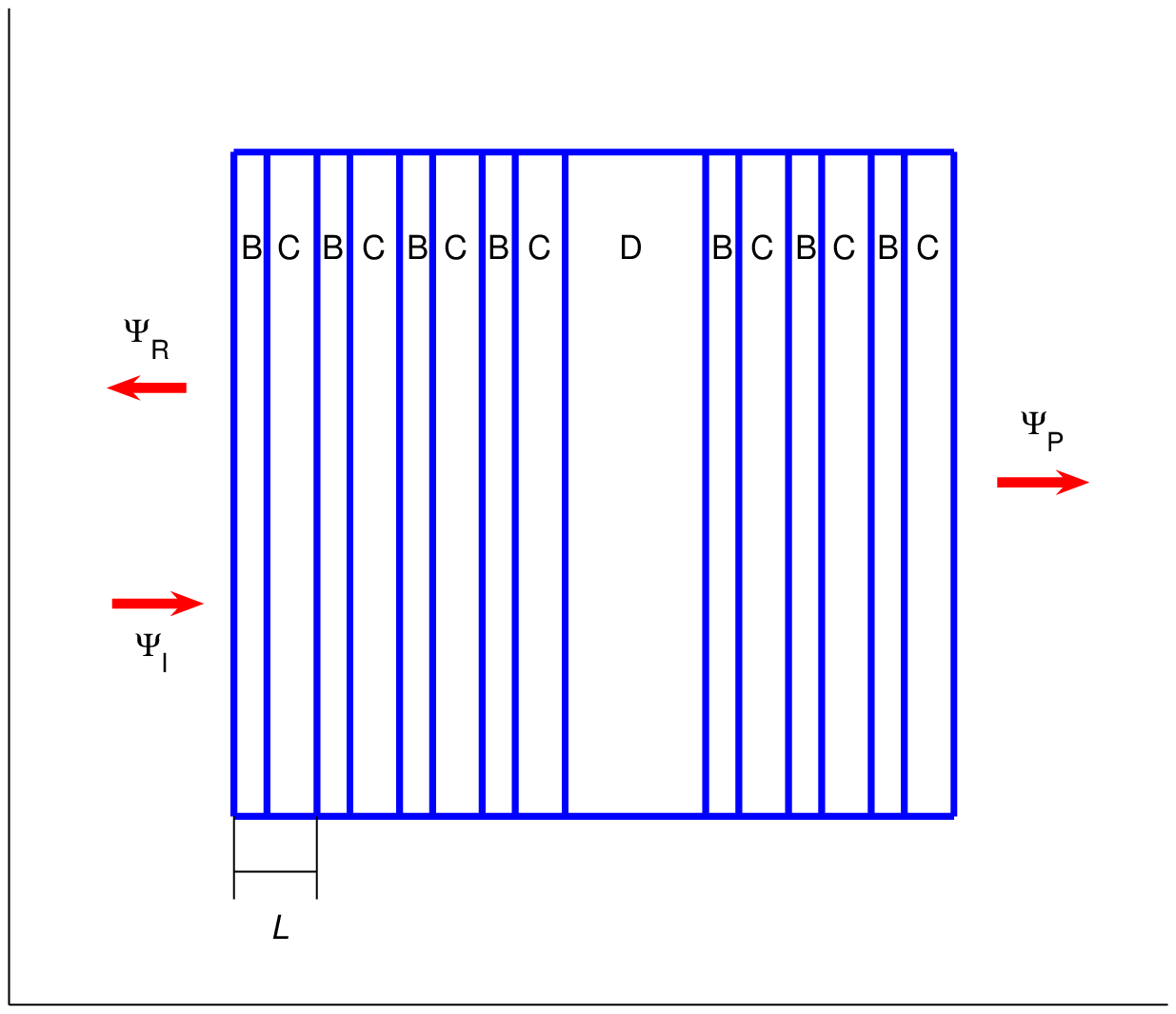}}
\caption{(Color online) Magnetic resonance cavity composed of magnetic layer
D sandwiched between a pair of identical periodic non-magnetic stacks (Bragg
reflectors). The incident wave $\Psi_{I}$ is linearly polarized with $%
E\parallel x$. Due to the nonreciprocal circular birefringence of the
magnetic material of D-layer, the reflected wave $\Psi_{R}$ and the
transmitted wave $\Psi_{P}$ are both elliptically polarized.}
\label{MC}
\end{figure}

\begin{figure}[tbph]
\scalebox{0.8}{\includegraphics[viewport=-50 0 500 300,clip]{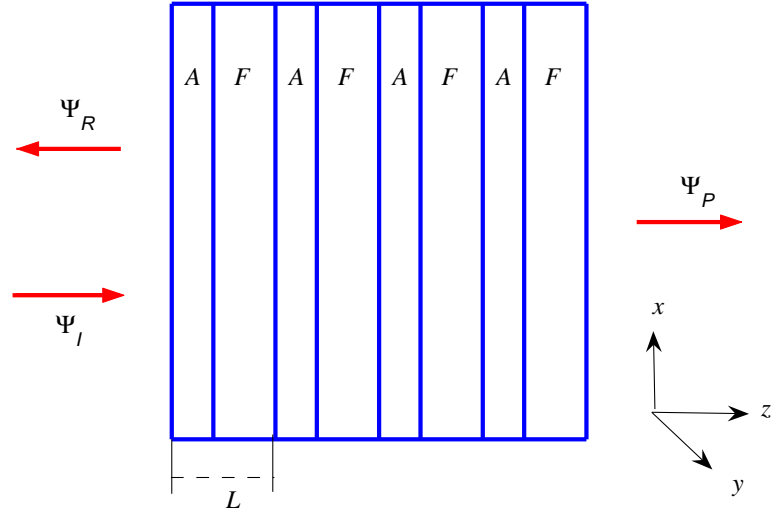}}
\caption{(Color online) Periodic layered structure composed of alternate
magnetic (F) and dielectric (A) layers. The F-layers are made of the same
lossy magnetic material as the D-layer in Fig. \protect\ref{MC}. $L$ is the
unit cell length. The incident wave $\Psi_{I}$ is linearly polarized with $%
E\parallel x$. Due to the nonreciprocal circular birefringence of the
magnetic material of the F-layers, the reflected wave $\Psi_{R}$ and the
transmitted wave $\Psi_{P}$ are both elliptically polarized.}
\label{AFStack}
\end{figure}

\begin{figure}[tbph]
\scalebox{0.8}{\includegraphics[viewport=-50 0 500
300,clip]{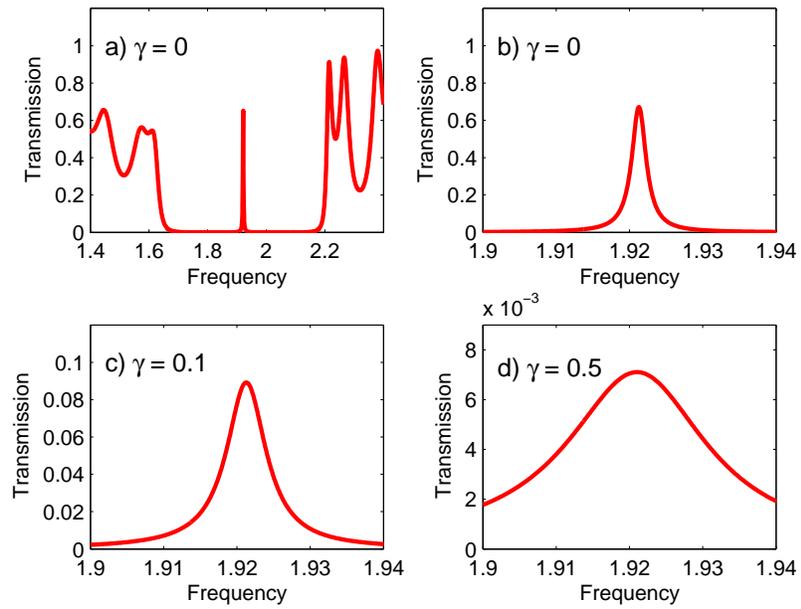}}
\caption{(Color online) Transmission dispersion of the layered array in Fig. 
\protect\ref{MC} for different values of absorption coefficient $\protect%
\gamma$ of the D-layer. Circular birefringence $\protect\alpha$ is
negligible. Fig. (b) shows the enlarged portion of Fig. (a) covering the
vicinity of microcavity resonance.}
\label{t_FSDM_A0_G}
\end{figure}

\begin{figure}[tbph]
\scalebox{0.8}{\includegraphics[viewport=-50 0 500
300,clip]{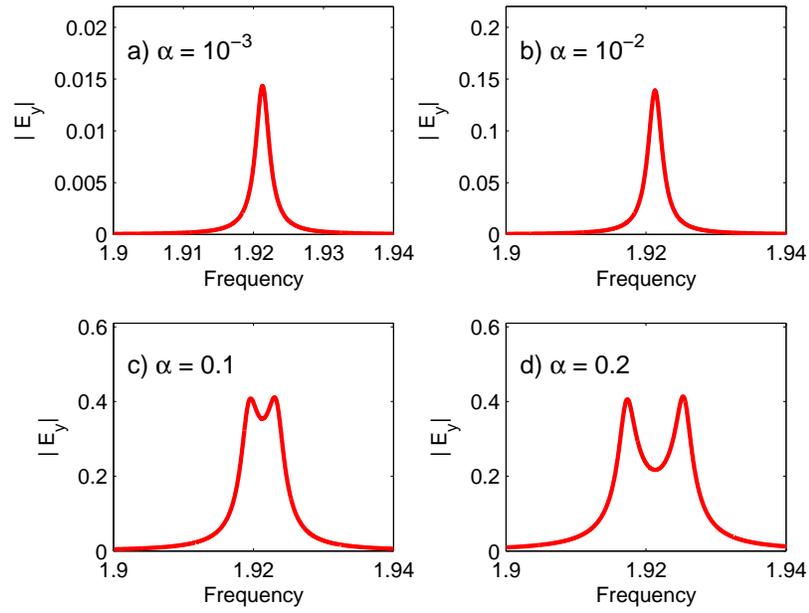}}
\caption{(Color online) Frequency dependence of polarization component $%
\left\vert E_{y}\right\vert $ of the wave transmitted through layered array
in Fig. \protect\ref{MC} for different values of circular birefringence $%
\protect\alpha$ of the D-layer and zero absorption. When circular
birefringence $\protect\alpha$ is strong enough, the cavity resonance splits
into a pair of twin resonances, corresponding to two circularly polarized
modes with opposite sense of rotation. The incident wave is linearly
polarized with $\vec{E}\parallel x$.}
\label{p_FSDM_A_G0}
\end{figure}

\begin{figure}[tbph]
\scalebox{0.8}{\includegraphics[viewport=-50 0 500 300,clip]{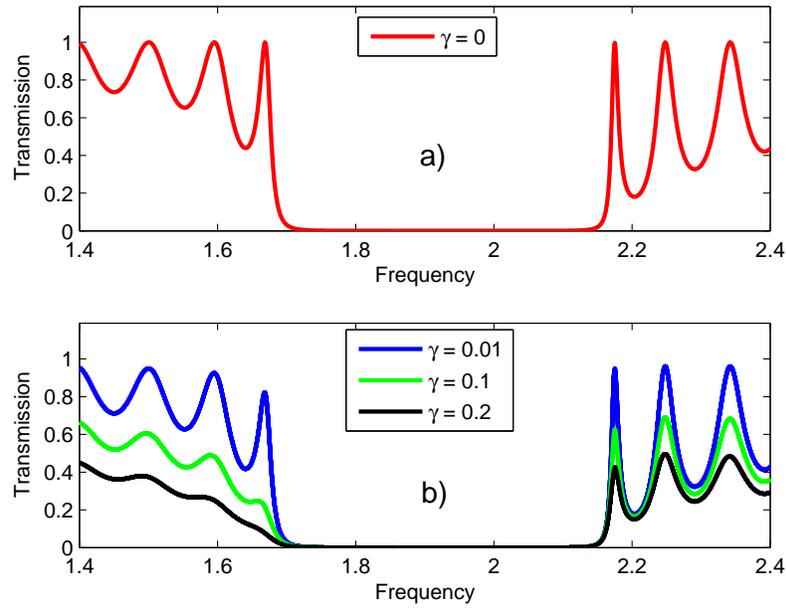}}
\caption{(Color online) Transmission dispersion of periodic layered
structure in Fig. \protect\ref{AFStack} for different values of absorption
coefficient $\protect\gamma$ of the F-layers. Circular birefringence $%
\protect\alpha$ is negligible.}
\label{t_PFS_a0_g}
\end{figure}

\begin{figure}[tbph]
\scalebox{0.8}{\includegraphics[viewport=-50 0 500
300,clip]{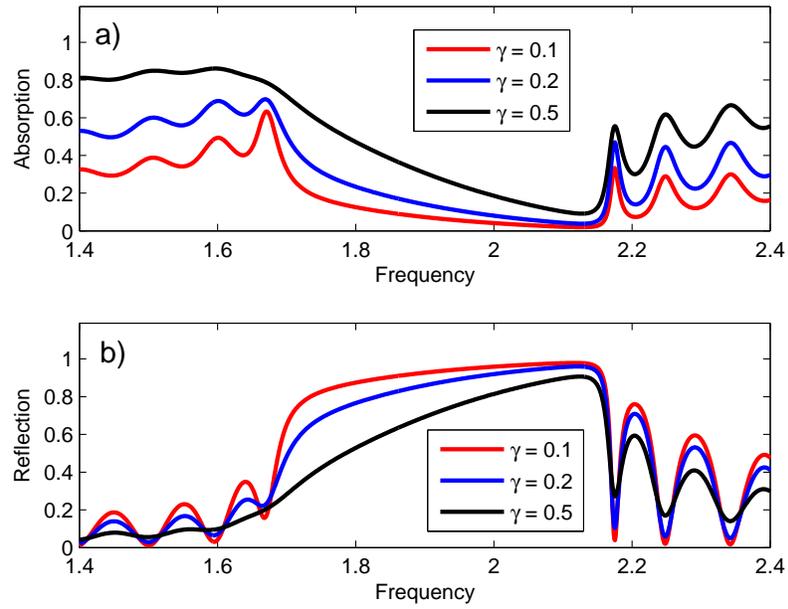}}
\caption{(Color online) Frequency dependence of (a) absorption and (b)
transmission of periodic layered structure in Fig. \protect\ref{AFStack} for
different values of absorption coefficient $\protect\gamma$ of the F-layers.
Circular birefringence $\protect\alpha$ is negligible.}
\label{ar_PFS_a0_g}
\end{figure}

\begin{figure}[tbph]
\scalebox{0.8}{\includegraphics[viewport=-50 0 500 300,clip]{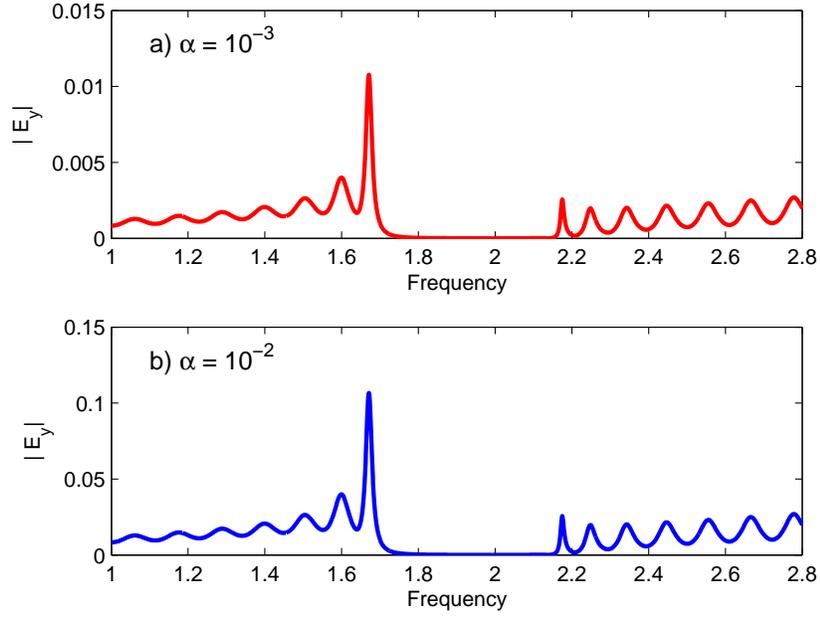}}
\caption{(Color online) Frequency dependence of polarization component $%
\left\vert E_{y}\right\vert $ of the wave transmitted through the periodic
layered structure in Fig. \protect\ref{AFStack} for different values of
circular birefringence $\protect\alpha$ of the F-layers and zero absorption.
The incident wave is linearly polarized with $\vec{E}\parallel x$.}
\label{p_PFS_a_g0}
\end{figure}

\begin{figure}[tbph]
\scalebox{0.8}{\includegraphics[viewport=-50 0 500
300,clip]{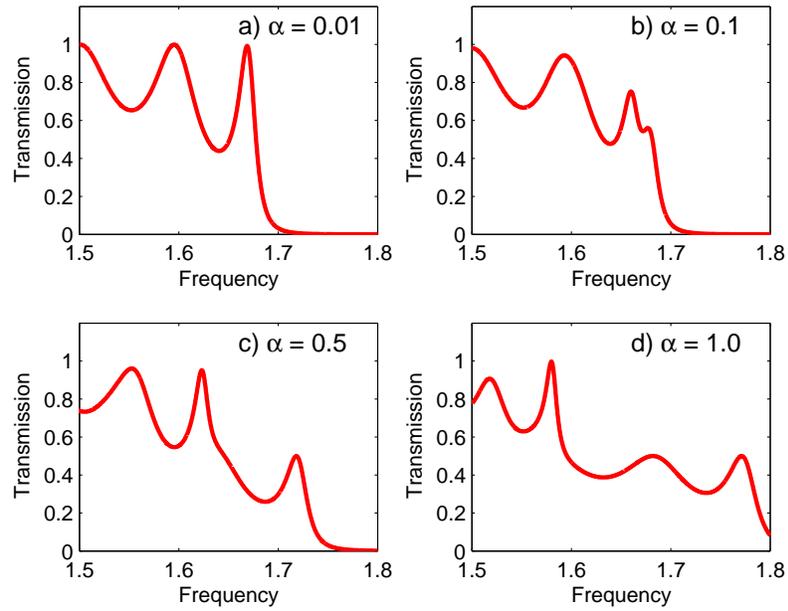}}
\caption{(Color online) Transmission dispersion of periodic layered
structure in Fig. \protect\ref{AFStack} for different values of circular
birefringence $\protect\alpha$ of the F-layers and zero absorption. When
circular birefringence $\protect\alpha$ is large enough, each transmission
resonance splits into a pair of twin resonances, corresponding to two
circularly polarized modes with opposite sense of rotation. The incident
wave polarization is linear.}
\label{t_PFS_a_g0_fwLx}
\end{figure}

\begin{figure}[tbph]
\scalebox{0.8}{\includegraphics[viewport=-50 0 500
300,clip]{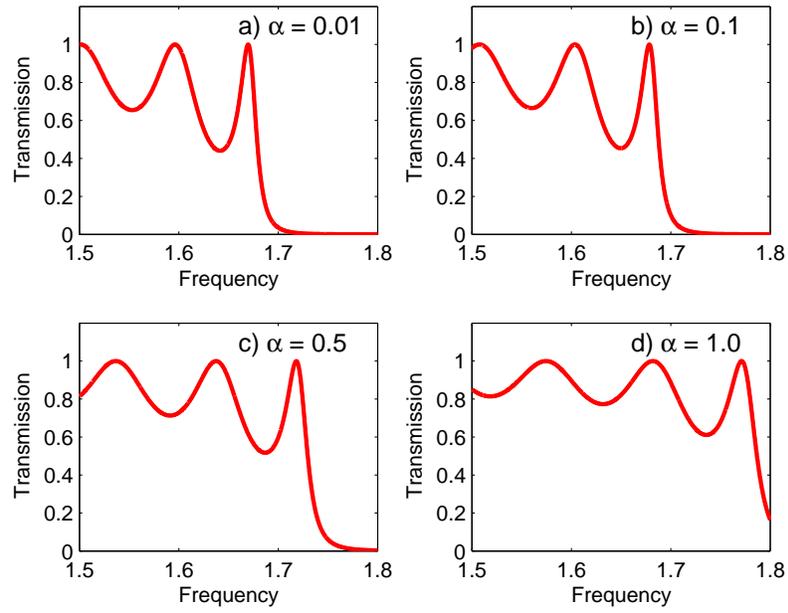}}
\caption{(Color online) The same as in Fig. \protect\ref{t_PFS_a_g0_fwLx},
but the incident wave polarization is circular with positive sense of
rotation.}
\label{t_PFS_a_g0_fwLcp}
\end{figure}

\begin{figure}[tbph]
\scalebox{0.8}{\includegraphics[viewport=-50 0 500
300,clip]{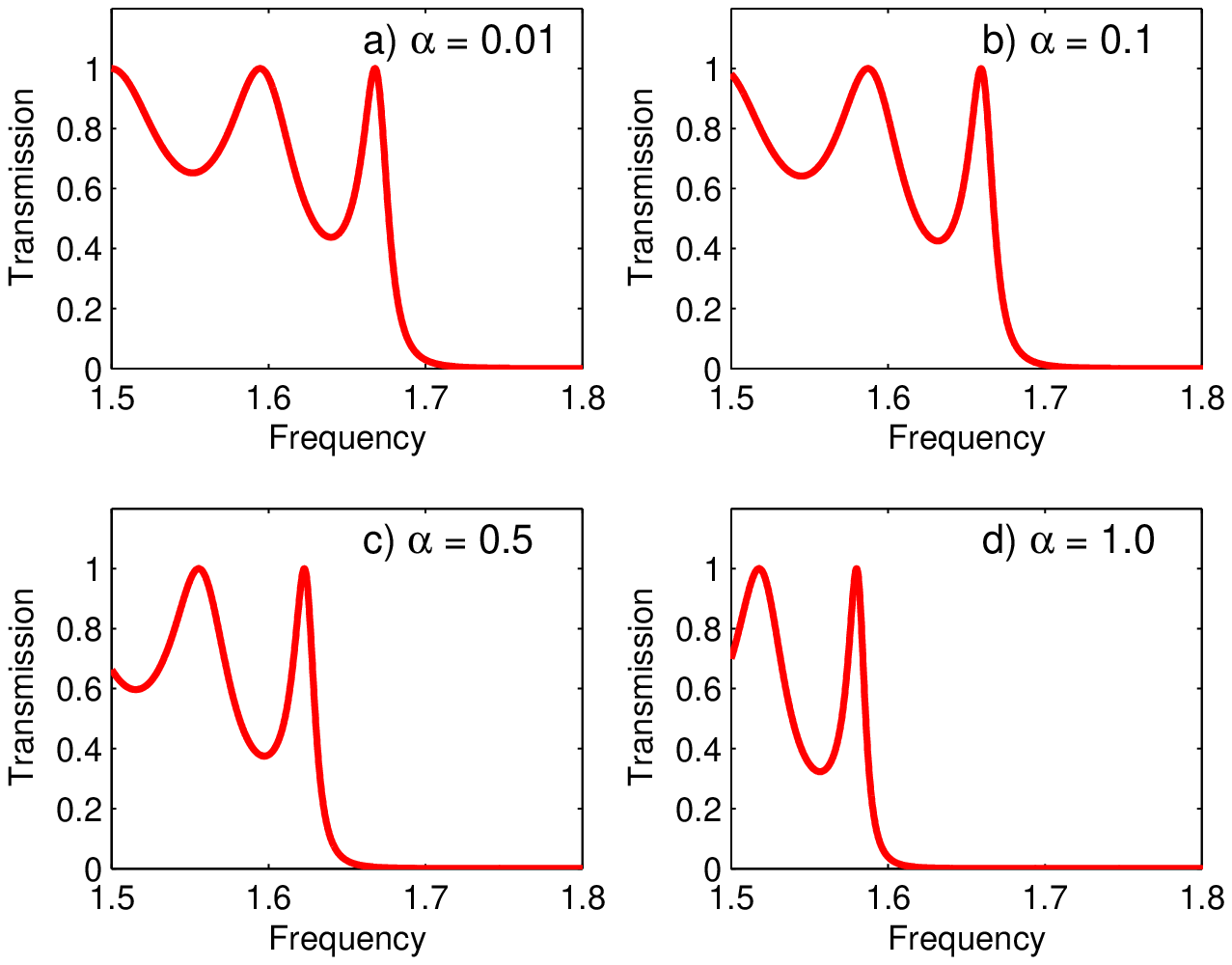}}
\caption{(Color online) The same as in Figs. \protect\ref{t_PFS_a_g0_fwLx}
and \protect\ref{t_PFS_a_g0_fwLcp}, but the incident wave polarization is
circular with negative sense of rotation.}
\label{t_PFS_a_g0_fwLcn}
\end{figure}

In the case of magnetic cavity resonance, the situation is fundamentally
different. In this case, the energy losses associated with absorption cannot
be much different from those of slow wave resonator, provided that both
arrays display comparable enhancement of Faraday rotation. What is
fundamentally different is the reflectivity. An inherent problem with any
(localized) defect mode is that any significant absorption in defect layer
makes it inaccessible. Indeed, if the D-layer in Fig. \ref{MC} displays an
appreciable absorption, the entire structure becomes highly reflective. As a
consequence, a major portion of the incident light energy is reflected from
the stack surface and never even reaches the magnetic D-layer. Such a
behavior is illustrated in Fig. \ref{r_FSDM_A0_G}, where we can see that as
soon as the absorption coefficient $\gamma $ exceeds certain value, further
increase in $\gamma $ leads to high reflectivity of the layered structure.
In the process, the total absorption $a$ reduces, as seen in Fig. \ref%
{a_FSDM_A0_G}, but the reason for this reduction is that the light simply
cannot reach the magnetic layer. There is no Faraday rotation enhancement in
this case.

\textbf{Acknowledgments:} Effort of A. Figotin and I. Vitebskiy is sponsored
by the Air Force Office of Scientific Research, Air Force Materials Command,
USAF, under grant number FA9550-04-1-0359.

\bigskip\pagebreak

\pagebreak

\pagebreak

\pagebreak

\pagebreak

\pagebreak

\begin{figure}[tbph]
\scalebox{0.8}{\includegraphics[viewport=-50 0 500
300,clip]{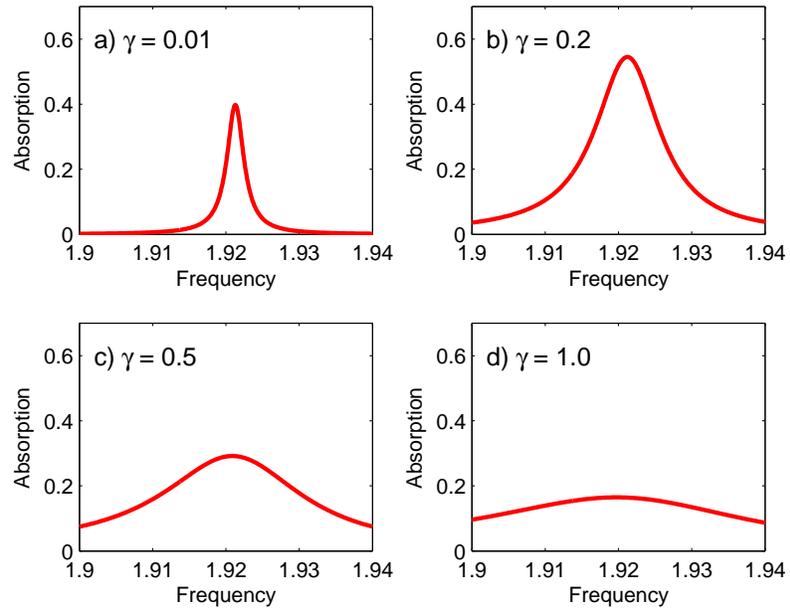}}
\caption{(Color online) Frequency dependence of absorption of the layered
array in Fig. \protect\ref{MC} for different values of absorption
coefficient $\protect\gamma$ of the D-layer. Circular birefringence $\protect%
\alpha$ is negligible. The frequency range shown covers the vicinity of
microcavity resonance. Observe that the stack absorption decreases after
coefficient $\protect\gamma$ exceeds certain value, which is in sharp
contrast with the case of a periodic stack, shown in Figs. \protect\ref%
{ar_PFS_a0_g}.}
\label{a_FSDM_A0_G}
\end{figure}

\pagebreak

\begin{figure}[tbph]
\scalebox{0.8}{\includegraphics[viewport=-50 0 500
300,clip]{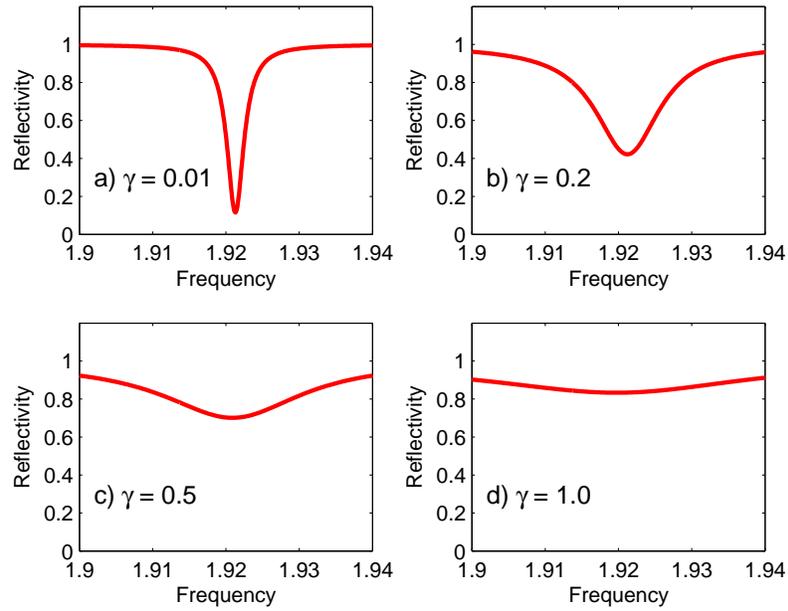}}
\caption{(Color online) Frequency dependence of the reflectance $r$ of the
layered array in Fig. \protect\ref{MC} for different values of absorption
coefficient $\protect\gamma$ of the D-layer. Circular birefringence $\protect%
\alpha$ is negligible. The frequency range shown covers the vicinity of
microcavity resonance. Observe that if the absorption coefficient $\protect%
\gamma$ of D-layer increases, the stack reflectivity also increases
approaching unity. Such a behaivior is line with frequency dependence of the
stack absorption shown in Fig. \protect\ref{a_FSDM_A0_G}. It is in sharp
contrast with the case of a periodic stack, shown in Figs. \protect\ref%
{ar_PFS_a0_g}.}
\label{r_FSDM_A0_G}
\end{figure}

\pagebreak

\pagebreak

\pagebreak

\pagebreak

\pagebreak

\pagebreak

\end{document}